# An updated review of (sub-)optimal diversification models

Johannes Bock

Department of Finance, Warwick Business School, University of Warwick, Scarman Rd, Coventry CV4 7AL, UK, email: j.bock@warwick.ac.uk

September 2018

# Abstract

In the past decade many researchers have proposed new optimal portfolio selection strategies to show that sophisticated diversification can outperform the naïve 1/N strategy in out-of-sample benchmarks. Providing an updated review of these models since DeMiguel *et al.* (2009b), I test sixteen strategies across six empirical datasets to see if indeed progress has been made. However, I find that none of the recently suggested strategies consistently outperforms the 1/N or minimum-variance approach in terms of Sharpe ratio, certainty-equivalent return or turnover. This suggests that simple diversification rules are not in fact inefficient, and gains promised by optimal portfolio choice remain unattainable out-of-sample due to large estimation errors in expected returns. Therefore, further research effort should be devoted to both improving estimation of expected returns, and possibly exploring diversification rules that do not require the estimation of expected returns directly, but also use other available information about the stock characteristics.

# Table of Contents

# List of figures

# List of tables

# Abbreviations

| | |
|---|---|
| CAPM | Capital Asset Pricing Model |
| CEQ | Certainty-equivalent return |
| GARCH | Generalized autoregressive conditional heteroskedasticity model |
| MV | Mean-variance optimal portfolio |
| SR | Sharpe ratio |
| VAR | Linear vector-autoregressive model of expected returns |
| VCV | Variance-covariance matrix |

# 1 Introduction and research question

Ever since Markowitz (1952) published his seminal portfolio selection paper, many researchers have tried to overcome the limitations of the classical mean-variance (*MV*) strategy. This is because it lacks stability and insufficiently accounts for estimation errors. Thus, small changes in the input parameters due to, for example, time-varying moments can cause large changes in the optimized portfolio. Hence, weights become sub-optimal and poor out-of-sample performance is observed. The error magnifying weight optimization based on unstable and noisy estimates of the moments has become widely known as the *error maximization problem* (Michaud, 1989). This is because during mean-variance optimization, securities with disproportionately large estimated returns and small variances are over-weighted, even though these may in fact be securities with large estimation errors.

Generally, as dictated by the mean-variance problem, there are two dimensions to this extensively researched issue, namely the efficient estimation (or forecasting) of the variance-covariance matrix and expected returns. This dissertation focuses on expected returns and compares recent optimal strategies to the naïve 1/N diversification and the popular minimum-variance approach. It contributes to existing literature by analysing whether any progress with newly proposed portfolio optimization strategies has been made since the study by DeMiguel *et al.* (2009b).

This research is motivated by empirical findings that the simple 1/N strategy outperforms the Markowitz and other sophisticated optimization rules in out-of-sample benchmarks (DeMiguel *et al.*, 2009b). Moreover, it is also a commonly used heuristic by investors who tend to irrationally ignore potential diversification benefits and simply spread their funds evenly across their chosen securities (Baltussen and Post, 2011; Benartzi and Thaler, 2001). Hence, it is especially interesting to see if the widely adopted naïve approach is in fact inefficient and whether there is any gain to adopting complex and estimation-heavy diversification rules. Additionally, there is a vast body of literature on portfolio selection strategies and the efficient estimation of moments. As of 2013 there are about 300 papers on expected return estimation alone (Green *et al.*, 2013). Especially, the accurate estimation of expected returns is crucial because MV-efficient portfolio weights are extremely sensitive to it (Best and Grauer, 1991). Therefore, in this paper the focus is not on efficient estimators of the variance-covariance of returns. This dissertation is not only practically relevant but also a contribution to existing research by providing an updated review of the effectiveness of weight optimization strategies put forth in the past decade.

The remaining structure of this dissertation is as follows: First, I will critically review relevant literature on portfolio selection strategies (section 2). Next, I describe the empirical datasets, provide definitions for all portfolio selection models considered and outline the evaluation methodology including metrics and statistical testing procedures (section 3). Finally, I discuss my empirical findings, limitations (section 4) and conclude with a summary of my results and their implications (section 5).

## 2   Literature review

The context of this dissertation is a large body of literature on sophisticated asset allocation models that go beyond the classical Markowitz strategy. It can broadly be categorized in Bayesian and non-Bayesian approaches to reducing moment estimation errors.

*Bayesian Approach*

The Bayesian approach is a popular choice, since it provides a natural solution to the parameter uncertainty issue by automatically accounting for estimation errors. The literature mainly differs on two aspects: how to choose an appropriate prior, and whether to base it on statistical or economic considerations.

**Diffuse priors.** This category of models is purely based on a statistical approach and relies on diffuse priors (Barry, 1974). However, their performance tends to be poor and statistically indistinguishable from the classical mean-variance strategy, since they do not use any prior information about the parameters (Tu and Zhou, 2010, p.963; DeMiguel *et al.*, 2009b, p.1917).

**Shrinkage.** Primarily, to reduce extreme estimation errors in expected returns, shrinkage introduces a small bias in the estimates and is attributed to James and Stein (1961). In early versions, sample means are 'shrunk' towards the common grand mean, which reduces the variance of estimates (Jobson and Korkie, 1980; James and Stein, 1961). A popular choice is the estimator introduced by Jorion (1986), as it also accounts for covariances and has thus been widely adopted (Tu and Zhou, 2011; DeMiguel *et al.*, 2009b). More recently, other approaches propose to also shrink the elements of the variance-covariance (VCV) matrix towards a single-factor model (Ledoit and Wolf, 2003) or the constant correlation estimate (Ledoit and Wolf, 2004a; 2004b). However, as shown by Jagannathan and Ma (2003), this is equivalent to imposing short-sale constraints on the minimum-variance portfolio. On the other hand, this may not necessarily hold for non-linear shrinkage methods, which shrink each of the sample VCV-matrix eigenvalues individually and determine the optimal shrinkage intensity for each based on its magnitude (Ledoit and Wolf, 2017).

**Economic priors.** Another Bayesian approach is to form an informative prior that is based on some reasonable economic belief to reduce the arbitrariness of the statistical methods. Under Pástor (2000) and Pástor and Stambaugh (2000) the shrinkage factor and target are based on the investor's belief in an asset-pricing model (*i.e.* a prior belief about the asset's mispricing $\alpha$) such as the Fama-French three-factor model. More generally, priors can incorporate any economic objective of investors, who may have some *ex ante* idea for the range of weights to be allocated to a security, for instance (Tu and Zhou, 2010).

*Non-Bayesian Approach*

A broader set of models can be summarized as non-Bayesian approaches, where not only extensions of the classical framework, but also complete departures from it, are proposed.

**Moment restrictions.** To reduce estimation error a restriction on the moment estimates can be imposed. This is achieved by reducing the Markowitz problem and only focusing on a single moment, or by exploiting the restrictions implied by the factor structure of returns. For instance, one can minimize portfolio variance alone. On the other hand, one can use MacKinlay and Pástor's (2000) 'missing-factor' model, which exploits mispricings embedded in the residuals covariance-matrix to improve expected return estimates.

**Robust optimization.** The robust formulation of the mean–variance model focuses on an investor that chooses the best portfolio under the worst possible scenario. The worst prior is chosen from an uncertainty set, which can be related to expected returns, covariances or some prior rule, for example (Anderson and Cheng, 2016, p.1362; Goldfarb and Iyengar, 2003, p.5; Lorenzo *et al.*, 2007, p.42-43). This is based on the multiple-priors and ambiguity-aversion literature coined by Gilboa and Schmeidler (1989) and Epstein and Schneider (2003). However, the value of robust portfolio models is unclear, as their performance seems to lie between the sample MV and the minimum-variance portfolio (DeMiguel *et al.*, 2009b, p.1927).

**Moments forecasting.** Using daily data DeMiguel *et al.* (2014) model stock returns and exploit serial correlations by fitting vector-autoregressive models, for example. However, this approach may suffer from the fact that while sampling error is reduced, it is at the cost of model specification error (Martellini and Ziemann, 2010, p.1468). The same issue applies to forecasting covariances using constant covariance models or dynamic multivariate GARCH-models (Engle, 2002; Chan *et al.*, 1999). To overcome this issue, DeMiguel *et al.* (2013) suggest the use of forward-looking option implied volatilities and correlations instead.

**Mean-variance timing.** To reduce turnover and improve over naïve diversification, Kirby and Ostdiek (2012) propose two simple timing strategies, which rebalance based on (1) estimated changes in conditional volatilities, and (2) reward-to-risk ratios. Similarly, Moreira and Muir (2017) scale returns by the inverse of their previous month's realized variance, which leads to less risk exposure when variance was recently high and vice versa. They find that their volatility-managed portfolios increase Sharpe ratios and produce utility gains for a mean-variance investor.

**Portfolio constraints.** Further, both Jagannathan and Ma (2003) and DeMiguel *et al.* (2009b) show that short-sale constraints improve performance and argue that it is as efficient as linear shrinkage. Moreover, DeMiguel *et al.* (2009a) suggest to impose additional norm constraints, which require the norm of the vector of portfolio weights to be smaller than a given threshold. Other researchers propose variants of the MV-portfolio, with diversification constraints on the weights of its components such as the 'equally-weighted risk contribution' or 'maximum diversification' portfolio (Maillard *et al.*, 2010; Choueifaty and Coignard, 2008).

**Firm characteristics.** Since many extensions of the MV strategy exhibit weak out-of-sample properties, Brandt *et al.* (2009) suggest a novel approach of forming portfolios by modelling the equity portfolio weights directly using asset characteristics as predictive variables. By maximizing average investor utility over a historic sample period given some stock characteristics, such as capitalization, book-to-market ratio and lagged return, Brandt *et al.* (2009) find robust out-of-sample performance.

**Higher-order moments.** In the presence of non-normality of returns, Martellini and Ziemann (2010) suggest to make the mean-variance problem four-dimensional including expected returns, second, third and fourth moments of asset returns.[1] By reducing the dimensionality of the higher-order moments estimation problem, they document superior out-of-sample performance over simple mean-variance optimization. However, modelling higher-order moments is still challenging in practice, due to an exponential increase in parameters (Brandt *et al.*, 2009).

**Optimal combination of portfolios.** Finally, similarly to shrinkage, the combination of portfolios is used to balance the trade-off between bias and variance. For example, 1/N is a highly biased strategy with no variance, whereas sophisticated optimization schemes are often unbiased asymptotically, but exhibit high variance in small samples (Tu and Zhou, 2011, p.2). By Bayesian-averaging or optimizing combination weights for a mean-variance investor, past research documents out-of-sample performance gains through combining high bias with high variance strategies (Anderson and Cheng, 2016; Tu and Zhou, 2011; DeMiguel *et al.*, 2009b; Kan and Zhou, 2007).

To conclude, recently proposed portfolio selection models that focus on reducing the impact of estimation error of expected returns considered in this study are summarized in Table 1.

*Performance evaluation of optimal portfolio strategies*

Aside from proposing new optimization strategies, researchers are also concerned with evaluating the performance of the various asset allocation strategies using simulation or empirical studies. While DeMiguel *et al.* (2009b) find empirically that optimal strategies fail to outperform naïve diversification consistently, other studies report the opposite result (Kirby and Ostdiek, 2012; Tu and Zhou, 2011). These findings are largely based on rolling-window procedures to estimate out-of-sample Sharpe ratios, certainty equivalents and turnover for the investigated strategies.

---

[1] Modelling higher-order moments is not only important in the presence of non-normally distributed returns. More generally, all risk diversification approaches adopting Markowitz preferences are subject to criticism on the basis of failing to use an appropriate measure for risk, which has traditionally been the second moment of returns. This is because research shows that investor preferences go beyond simple volatility measures as they deliberately under-diversify from a mean-variance perspective. For instance, investors derive utility from positively skewed asset return characteristics (Mitton and Vorkink, 2007). This has also been noted by Kahneman and Tversky (1992; 1979), in their seminal papers on prospect theory and asymmetric preferences in the domain of highly and less probable losses and gains (*i.e.* fourfold pattern of risk attitudes).

**Table 1 | Portfolio selection models considered**

The table lists the various portfolio selection models considered. The last two columns give the abbreviation used to refer to the strategy and a reference to past research where it was proposed. Note that some of these optimization rules have already been studied by DeMiguel *et al.* (2009b), however since they are required for recently proposed combination strategies, they are still reported with updated data.

| # | Model | Abbreviation | Reference |
|---|---|---|---|
| **Naïve** | | | |
| 0 | 1/N (*benchmark strategy*) | ew | |
| **Classical approach** | | | |
| 1 | Sample mean-variance | mv | |
| **Shrinkage estimators** | | | |
| 2 | Bayes-Stein | bs | Jorion (1986) |
| **Moment forecasting** | | | |
| 3 | Conditional mean-variance with vector auto-regressive model | mv-var | DeMiguel *et al.* (2014) |
| **Moment restrictions** | | | |
| 4 | Minimum-variance | min | |
| **Mean-variance timing** | | | |
| 5 | Reward-to-risk-timing | rrt | Kirby and Ostdiek (2012) & Moreira and Muir (2017) |
| **Robust portfolio choice** | | | |
| 6 | Robust portfolios | ac-mv, ac-bs, ac-mv-min | Anderson and Cheng (2016) |
| **Optimal combination of portfolios** | | | |
| 7 | 'Three-fund' model | mv-min | Kan and Zhou (2007) |
| 8 | Mixture of 1/N and minimum-variance | ew-min | DeMiguel *et al.* (2009b) |
| 9 | Mixture of 1/N and mean-variance | ew-mv | Tu and Zhou (2011) |
| 10 | Mixture of 1/N and 'three fund' model (Kan and Zhou, 2007) | ew-mv-min | Tu and Zhou (2011) |
| **Portfolio constraints** | | | |
| 11 | Sample mean-variance with short-sale constraints | mv-c | |
| 12 | Minimum-variance with short-sale constraints | min-c | |
| 13 | Bayes-Stein with short-sale constraints | bs-c | |
| 14 | Robust portfolios with short-sale constraints | ac-mv-c, ac-bs-c, ac-mv-min-c | |
| 15 | Norm-constrained minimum-variance | min-norm | DeMiguel *et al.* (2009a) |
| 16 | Norm-constrained Bayes-Stein | bs-norm | DeMiguel *et al.* (2009a) |

# 3    Methodology and data

The following section describes the six empirical datasets, provides definitions for all portfolio selection models and outlines the evaluation metrics and statistical testing procedures.

## 3.1    Data description and summary

Table 2 summarizes the datasets used in this study which is guided by previous research. However, past literature is being extended by including stock level data due to criticism of considering only portfolio data subject to high estimation risk and extreme turnover (Kirby and Ostdiek, 2012).

**Table 2 | Datasets considered**

The table lists the various datasets analysed, the number of risky assets *N*, where the number after the '+' indicates the number of factor portfolios available and the time-period for which the data is available. Each dataset contains monthly excess returns over the one-month US T-Bill return (from Kenneth French Data Library). To avoid distortions and survivorship bias, note that for datasets 5 and 6 all constituents which have historically been in the index for at least one day and have at least 241 observations are included. For a more detailed description of the datasets, see Appendix A.

| # | Dataset | N | Time-period | Source | Abbreviation |
|---|---------|---|-------------|--------|--------------|
| 1 | 10 industry portfolios and US equity market portfolio (MKT) | 10+1 | 06/1963 – 03/2018 | Kenneth French Data Library | Industry |
| 2 | Eight MSCI country indices and the World index | 8+1 | 02/1995 – 03/2018 | Bloomberg | International |
| 3 | SMB, HML, UMD portfolios and MKT | 3+1 | 06/1963 – 03/2018 | Kenneth French Data Library | SMB/HML/UMD |
| 4 | 25 portfolios formed on size-and book-to-market and the MKT, SMB, HML and UMD portfolios | 25+4 | 06/1963 – 03/2018 | Kenneth French Data Library | FF-4 |
| 5 | DJIA constituents and the MKT, SMB, HML and UMD portfolios | 42+4 | 06/1963 – 03/2018 | Compustat/CRSP | DJIA |
| 6 | S&P 500 index constituents and the MKT, SMB, HML and UMD portfolios | 938+4 | 06/1963 – 03/2018 | Compustat/CRSP | SP500 |

Moreover, according to a simulation study by DeMiguel *et al.* (2009b, p.1920), on the one hand, the 1/N strategy is likely to outperform optimal diversification schemes when the number of parameters is large (*i.e.* large number of risky assets *N*), but on the other hand, fails to dominate if levels of idiosyncratic volatility are high. As individual stocks exhibit a more significant idiosyncratic component in comparison to portfolio data, datasets 5 and 6 are included to investigate this trade-off empirically. Two datasets with stock level returns were included to ensure robustness and analyse performance for different levels of *N*. Further, it is important to note that the number of available assets may vary at each investment date *t* when the portfolio is being constructed, since for datasets 5 and 6 not all risky assets contain observations for the full time-period.

Moreover, even though for dataset 6 there are a total of 500 assets available at each date *t*, following DeMiguel *et al.* (2014, p.1035), I only randomly sample 60 assets from the available security universe at each date *t* instead of investing in all of them to avoid small sample issues. This is because I use sample estimates of the VCV-matrix, which would become problematic if the number of stocks becomes the same order of magnitude as the number of observations per stock (Ledoit and Wolf, 2003, p.604). I discuss further details of the estimation procedure in section 3.2 and 3.3.

As summarized in Table 4 in Appendix B, all empirical datasets contain highly leptokurtotic and (positively) skewed returns data. However, this is especially true for datasets 5 and 6 where some risky assets exhibit kurtosis and skewness values of up to 117.23 and 8.47, respectively. Moreover, according to the Ljung–Box test statistics reported in Table 5 in Appendix B, I find significant serial correlation at the 1% confidence level for some risky assets. This is important to note as I report non-robust test statistics. I discuss further implications of this for my results in section 4.

## 3.2 Description of asset allocation models considered

In this section I describe the portfolio selection models summarized in Table 1. I denote $R_t$ the *N*-vector of excess returns over the risk-free asset, where *N* denotes the number of risky assets available at time *t*. Accordingly, $\mu_t$ and $\Sigma_t$ are the expected excess returns and the *NxN* variance-covariance matrix at time *t* with their sample counterparts $\hat{\mu}_t$ and $\hat{\Sigma}_t$, respectively. I denote *M* the length of the moments estimation window and *T* the total number of observations available. $\mathbf{1}_N$ denotes a *N*-dimensional vector of ones. Furthermore, $x_{k,t}$ is a vector of portfolio weights invested in the risky assets and $1 - \mathbf{1}'_N x_{k,t}$ invested in the risk-free asset using strategy *k*. Finally, $w_{k,t}$ is the vector of *relative* weights invested in *N* risky assets at each investment date *t*.

### 3.2.1 Naïve portfolio

The 1/N benchmark strategy (*ew*) simply invests equal amounts across assets completely disregarding any further information about the asset's characteristics. It invests a share of

$$w_{i,ew,t} = \frac{1}{N} \qquad (1)$$

in each risky asset *i*.

### 3.2.2 Sample mean-variance portfolio

The classical sample-based mean-variance approach (*mv*) maximizes Markowitz investor utility with a coefficient of risk aversion $\gamma$

$$\max_{x_t} \; x_t' \mu_t - \frac{\gamma}{2} \, x_t' \Sigma_t \, x_t \qquad (2)$$

where the optimal solution is $x_{mv,t} = \frac{1}{\gamma} \Sigma_t^{-1} \mu_t$ and relative portfolio weights are given by

$$w_{mv,t} = \frac{\Sigma_t^{-1} \mu_t}{\mathbf{1}_N' \Sigma_t^{-1} \mu_t}. \qquad (3)$$

The sample estimates $\hat{\mu}_t$ and $\hat{\Sigma}_t$ are computed by using the most recent *M* observations

$$\hat{\mu}_t = \frac{1}{M} \sum_{s=t-M+1}^{t} R_s \qquad (4)$$

$$\hat{\Sigma}_t = \frac{1}{M-1} \sum_{s=t-M+1}^{t} (R_s - \hat{\mu}_t)(R_s - \hat{\mu}_t)'. \qquad (5)$$

Finally, $\hat{\mu}_t$ and $\hat{\Sigma}_t$ are inserted into Eq. 3. Note that I will use the sample VCV-estimate $\hat{\Sigma}_t$ for all following models to ensure comparability, even though past research may use different estimators of second moments.

### 3.2.3 Bayes-Stein shrinkage portfolio

To deal with the error in estimating expected returns, the Bayes-Stein shrinkage (*bs*) portfolio uses a Bayesian estimate of $\mu_t$ based on a *subjective prior*. I implement the Bayes-Stein estimator introduced by Jorion (1986) of the form

$$\hat{\mu}_{bs,t} = (1 - \hat{\phi}_t) \hat{\mu}_t + \hat{\phi}_t \hat{\mu}_{min,t}, \qquad where \qquad (6)$$

$$\hat{\phi}_t = \frac{N+2}{(N+2) + M(\hat{\mu}_t - \hat{\mu}_{min,t})' \hat{\Sigma}_t^{-1}(\hat{\mu}_t - \hat{\mu}_{min,t})} \qquad (7)$$

$$\hat{\mu}_{min,t} = \hat{\mu}_t' \hat{w}_{min,t} \qquad (8)$$

which 'shrinks' the sample mean toward the common 'grand mean', that is the mean of the minimum-variance portfolio $\mu_{min}$ (as defined in section 3.2.5). The quantities $\hat{\mu}_{bs,t}$ and $\hat{\Sigma}_t$ are thereupon inserted into Eq. 3 to get $\hat{w}_{bs,t}$.[2]

---

[2] Note that Jorion suggests using a scaled VCV-estimate $\frac{M-1}{M-N-2} \hat{\Sigma}_t$ instead.

### 3.2.4 Conditional mean-variance with vector autoregressive model

The conditional mean-variance approach, with a vector-autoregressive model (*mv-var*) proposed by DeMiguel *et al.* (2014), is based on a norm-constrained mean-variance portfolio of the form

$$\min_{w_t} \ w_t' \Sigma_t w_t - \frac{1}{\gamma} w_t' R_{t+1} \quad s.t. \tag{9}$$

$$w_t' \mathbf{1}_N = 1 \tag{10}$$

$$\|w_t - w_{min-c,t}\|_1 = \sum_{i=1}^{N} |w_{i,t} - w_{i,min-c,t}| \leq \delta \tag{11}$$

where $w_{min-c,t}$ represents the weights of the short-sale-constrained minimum-variance portfolio and $\delta$ is some threshold. Note that, unlike DeMiguel *et al.* (2014, p.1058), to ensure comparability across models I do not shrink the VCV-matrix at any point and use the sample estimate $\hat{\Sigma}_t$ instead. Like DeMiguel *et al.* (2014, p.1057-1058 & 1061) I will consider three threshold parameters: $\delta_1 = 2.5\%$, $\delta_2 = 5\%$ and $\delta_3 = 10\%$. These imply, in the case of $\delta_1 = 2.5\%$ for instance, that the sum of all negative weights in the norm-constrained conditional portfolios must be smaller than 2.5%. Moreover, returns follow a linear vector-autoregressive (VAR) model of the form

$$R_{t+1} = A + B \times R_t + \epsilon_{t+1} \tag{12}$$

where $A$ and $B$ are ridge-regression estimates with a L2-regularization parameter of $\alpha = 1$ and an error term $\epsilon_{t+1}$.

### 3.2.5 Minimum-variance portfolio

The minimum-variance (*min*) strategy completely ignores expected returns and thus, minimizes risk

$$\min_{w_t} \ x_t' \Sigma_t x_t \tag{13}$$

with optimal weights given by

$$w_{min,t} = \frac{\mathbf{1}_N' \Sigma_t^{-1}}{\mathbf{1}_N' \Sigma_t^{-1} \mathbf{1}_N} \tag{14}$$

in which the sample VCV-matrix estimate $\hat{\Sigma}_t$ is inserted.

### 3.2.6 Reward-to-risk timing portfolio

With their reward-to-risk timing (*rrt*) strategy Kirby and Ostdiek (2012) propose a simple long-only strategy based on the variance and factor loadings of the risky assets. Their idea is similar to Moreira and Muir (2017), who construct *volatility-managed portfolios* by scaling monthly returns by the inverse of their variance. To reduce estimation risk (*i.e.* to lower the sampling variation) of expected returns, Kirby and Ostdiek (2012) suggest to exploit the factor structure of returns and use the factor loadings $\beta_{ij,t}$ of the *i*th asset with respect to the *j*th factor in period *t* from a K-factor model. In contrast to their approach, I do not assume covariances to be zero to ensure comparability to other tested strategies. Hence, portfolio weights at time *t* are given by

$$\widehat{w}_{rrt,t} = \frac{\left(\Sigma_t^{-1}\bar{\beta}_t^+\right)^\omega}{\sum_{i=1}^N \left(\Sigma_t^{-1}\bar{\beta}_t^+\right)^\omega} \qquad (15)$$

where the tuning parameter $\omega$ measures the timing aggressiveness. Moreover, to enforce a long-only strategy Kirby and Ostdiek (2012) eliminate any asset *i* with negative average beta $\bar{\beta}_{i,t} < 0$ and therefore, $\bar{\beta}_{i,t}^+ = \max(0, \bar{\beta}_{i,t})$ and $\bar{\beta}_{i,t} = \frac{1}{K}\sum_{j=1}^K \beta_{ij,t}$. Depending on the empirical dataset the strategy is implemented using either the CAPM or the Carhart (1997) four-factor-model. Following Kirby and Ostdiek (2012), I consider $\omega_1 = 1$ and $\omega_2 = 2$.

### 3.2.7 Robust portfolios

Anderson and Cheng (2016, p.1349) suggest robust versions of the $\widehat{w}_{mv,t}$ (*ac-mv*), $\widehat{w}_{bs,t}$ (*ac-bs*) and $\widehat{w}_{mv-min,t}$ (*ac-mv-min*) portfolios, which allow uncertainty about future expectations. To decide on a portfolio that does well even under the worst possible scenario, the investor assumes first that the best approximation to the distribution of future excess returns is normal

$$R_{t+1} \sim N(\mu_{k,t}, \Sigma_t) \qquad (16)$$

where $\mu_{k,t}$ may vary across the three strategies. However, because the investor worries that her assumption is wrong, she solves the robust mean-variance optimization by

$$\min_{w_t}\left(-w_t'\mu_{k,t} - \frac{\log q_t}{2\tau} + \frac{\tau\, w_t'\Sigma_t w_t}{2q_t}\right), \quad where \qquad (17)$$

$$q_t = 1 - \gamma\, \tau\, w_t'\Sigma_t w_t, \quad s.t.\ \mathbf{1}_N' w_t = 1 \qquad (18)$$

in which $\tau$ is a measure of model uncertainty aversion and larger values correspond to higher levels of aversion. As $\tau \to 0$, Eq. 17 approaches the (non-robust) mean-variance problem similar to Eq. 2. Following Anderson and Cheng (2016, p.1351) I will consider a fixed model uncertainty parameter of $\tau = 4$. This is because, they show that their dynamic method is inferior to a fixed $\tau$ in the case of monthly data (Anderson and Cheng, 2016, p.1365).

### 3.2.8 Three-fund model

Mainly to improve models based on Bayes-Stein shrinkage estimators, Kan and Zhou (2007, p.636 & 643) suggest a 'three-fund' model (*mv-min*) which combines the sample mean-variance and minimum-variance portfolio weights optimally with respect to the expected utility of a mean-variance investor. Like DeMiguel *et al.* (2009b, p.1927), I compute relative weights by[3]

$$\widehat{w}_{mv-min,t} = \frac{\hat{x}_{mv-min,t}}{|\mathbf{1}'_N \hat{x}_{mv-min,t}|}, \quad where \tag{19}$$

$$\hat{x}_{mv-min,t} = \frac{c}{\gamma}\left[\left(\frac{\hat{\psi}_a^2}{\hat{\psi}_a^2 + \frac{N}{M}}\right)\hat{\Sigma}_t^{-1}\hat{\mu}_t + \left(\frac{\frac{N}{M}}{\hat{\psi}_a^2 + \frac{N}{M}}\right)\hat{\mu}_{g,t}\hat{\Sigma}_t^{-1}\mathbf{1}_N\right] \tag{20}$$

$$c = \frac{(M-N-1)(M-N-4)}{M(M-2)} \tag{21}$$

$$\hat{\mu}_{g,t} = \frac{\hat{\mu}'_t \hat{\Sigma}_t^{-1} \mathbf{1}_N}{\mathbf{1}'_N \hat{\Sigma}_t^{-1} \mathbf{1}_N} \tag{22}$$

$$\hat{\psi}^2 = (\hat{\mu}_t - \hat{\mu}_{g,t})' \hat{\Sigma}_t^{-1} (\hat{\mu}_t - \hat{\mu}_{g,t}) \tag{23}$$

$$\hat{\psi}_a^2 = \frac{(M-N-1) \times \hat{\psi}^2 - (N-1)}{M} + \frac{2 \times (\hat{\psi}^2)^{\frac{N-1}{2}}(1+\hat{\psi}^2)^{-\frac{M-2}{2}}}{M \times B_{\hat{\psi}^2/(1+\hat{\psi}^2)}\left(\frac{N-1}{2}, \frac{M-N+1}{2}\right)} \tag{24}$$

in which

$$B_z(a,b) = \int_0^z y^{a-1}(1-y)^{b-1} dy \tag{25}$$

is the incomplete beta function. Following Anderson and Cheng (2016, p.1346 &1349), the expected return of this strategy is computed by

$$\hat{\mu}_{mv-min,t} = (1-\xi_t)\hat{\mu}_t + \xi_t \hat{\mu}_{g,t}, \quad where \tag{26}$$

$$\xi_t = \frac{N}{M\hat{\psi}_a^2 + N} \tag{27}$$

---

[3] Kirby and Ostdiek (2012, p.446) point out that the rescaling of the weights for the 'three-fund' strategy effectively means to reduce it to a 'two-fund" model eliminating the risk-free asset.

and the second moments are again estimated using the sample VCV-matrix.

### 3.2.9 Optimal combination of naïve and optimized portfolios

The following diversification strategies propose an optimal combination of three sophisticated asset allocation rules with the naïve strategy to improve performance.

*Mixture with minimum-variance portfolio*

DeMiguel *et al.* (2009b, p.1927) suggest for the most part to ignore expected returns, due to their estimation difficulty, and combine the 1/N rule with the minimum-variance portfolio (*ew-min*). Specifically, they propose an optimal combination with regards to a MV-investor of the form

$$\widehat{w}_{ew-min,t} = \frac{\hat{x}_{ew-min,t}}{|\mathbf{1}'_N \hat{x}_{ew-min,t}|}, \ where \tag{28}$$

$$\hat{x}_{ew-min,t} = c \left[ \left( \frac{\hat{\psi}_a^2}{\hat{\psi}_a^2 + \frac{N}{M}} \right) \frac{1}{N} \mathbf{1}_N + \left( \frac{\frac{N}{M}}{\hat{\psi}_a^2 + \frac{N}{M}} \right) \hat{\mu}_{g,t} \hat{\Sigma}_t^{-1} \mathbf{1}_N \right] \tag{29}$$

in which the optimal combination weights are given by the Kan and Zhou (2007, p.643) three-fund separation rule.

*Mixture with sample mean-variance portfolio*

Moreover, Tu and Zhou (2011, p.28) look at various sophisticated portfolio strategies and find especially that combining naïve diversification with the $\widehat{w}_{mv-min,t}$ portfolio yields consistently positive and on average the highest certainty equivalent returns among the models they considered. Moreover, combining the 1/N with the MV portfolio yields the second largest certainty equivalent returns, and both combination strategies outperform the 1/N rule more often than their competing models, which is why both are considered here. Following Tu and Zhou (2011, p.5-6) the weights for the mixture with the mean-variance (*ew-mv*) portfolio are computed by

$$\widehat{w}_{ew-mv,t} = \frac{\hat{x}_{ew-mv,t}}{|\mathbf{1}'_N \hat{x}_{ew-mv,t}|}, \qquad where \tag{30}$$

$$\hat{x}_{ew-mv,t} = (1 - \hat{\lambda}_{ew-mv,t}) \times w_{ew,t} + \hat{\lambda}_{ew-mv,t} \times \frac{1}{\gamma} \hat{\Sigma}_t^{-1} \hat{\mu}_t \tag{31}$$

in which the optimal combination is given by

$$\hat{\lambda}_{ew-mv,t} = \frac{\hat{\pi}_1}{\hat{\pi}_1 + \hat{\pi}_2}, \qquad where \tag{32}$$

$$\hat{\pi}_1 = w'_{ew,t}\hat{\Sigma}_t w_{ew,t} - \frac{2}{\gamma}w'_{ew,t}\hat{\mu}_t + \frac{1}{\gamma^2}\hat{\theta}_a^2 \tag{33}$$

$$\hat{\pi}_2 = \frac{1}{\gamma^2}(h_1 - 1)\hat{\theta}_a^2 + \frac{h_1}{\gamma^2}\frac{N}{M} \tag{34}$$

$$\hat{\theta}^2 = \hat{\mu}'_t \hat{\Sigma}_t^{-1} \hat{\mu}_t \tag{35}$$

$$\hat{\theta}_a^2 = \frac{(M-N-2)\hat{\theta}^2 - N}{M} + \frac{2 \times \left(\hat{\theta}^2\right)^{\frac{N}{2}}(1+\hat{\theta}^2)^{-\frac{M-2}{2}}}{M \times B_{\hat{\theta}^2/(1+\hat{\theta}^2)}\left(\frac{N}{2}, \frac{M-N}{2}\right)} \tag{36}$$

$$h_1 = \frac{(M-2)(M-N-2)}{(M-N-1)(M-N-4)} \tag{37}$$

where $\hat{\theta}_a^2$ is defined by Kan and Zhou (2007, p.637 & 639).

*Mixture with 'three-fund' model*

Following the combination rule suggested by Tu and Zhou (2011, p.7) of the form

$$\widehat{w}_{ew-mv-min,t} = \frac{\hat{x}_{ew-mv-min,t}}{\left|\mathbf{1}'_N \hat{x}_{ew-mv-min,t}\right|}, \qquad where \tag{38}$$

$$\hat{x}_{ew-mv-min,t} = \left(1 - \hat{\lambda}_{ew-mv-min,t}\right) \times w_{ew,t} + \hat{\lambda}_{ew-mv-min,t} \times \hat{x}_{mv-min,t} \tag{39}$$

the optimal choice for the mixture with the three-fund model (*ew-mv-min*) is given by

$$\hat{\lambda}_{ew-mv-min,t} = \frac{\hat{\pi}_1 - \hat{\pi}_{13}}{\hat{\pi}_1 - 2\hat{\pi}_{13} + \hat{\pi}_3}, \qquad where \tag{40}$$

$$\hat{\pi}_{13} = \frac{1}{\gamma^2}\hat{\theta}_a^2 - \frac{1}{\gamma}w'_{ew,t}\hat{\mu}_t + \frac{1}{\gamma h_1}\left(\left[\hat{\psi}_a^2 w'_{ew,t}\hat{\mu}_t + (1-\hat{\psi}_a^2)\hat{\mu}_{g,t}w'_{ew,t}\mathbf{1}_N\right]\right.$$
$$\left. - \frac{1}{\gamma}\left[\hat{\psi}_a^2 \hat{\mu}'_t\hat{\Sigma}_t^{-1}\hat{\mu}_t + (1-\hat{\psi}_a^2)\hat{\mu}_{g,t}\hat{\mu}'_t\hat{\Sigma}_t^{-1}\mathbf{1}_N\right]\right) \tag{41}$$

$$\hat{\pi}_3 = \frac{1}{\gamma^2}\hat{\theta}_a^2 - \frac{1}{\gamma^2 h_1}\left(\hat{\theta}_a^2 - \frac{N}{M}\hat{\psi}_a^2\right) \tag{42}$$

in which $\hat{\psi}_a^2$ and $\hat{\mu}_{g,t}$ are the estimators of the squared slope of the asymptote to the minimum-variance frontier, and the expected excess return of the global minimum-variance portfolio as shown by Kan and Zhou (2007, p.643).

### 3.2.10 Constrained portfolios

*Short-sale constraints*

The sample mean-variance-constrained (*mv-c*), Bayes-Stein-constrained (*bs-c*), minimum-variance-constrained (*min-c*) and Robust-constrained (*ac-c*) strategies are considered by imposing additional nonnegativity constraints on the corresponding optimization schemes of the form $w_{k,t} \geq 0$.

*Norm-constraints*

Moreover, I consider norm-constrained versions of the minimum-variance (*min-norm*) and Bayes-Stein (*bs-norm*) portfolios as DeMiguel *et al.* (2009a) find that this type of constraint often yields a higher Sharpe ratio than the unconstrained counterpart. Portfolio weights are computed by additionally imposing the weight constraint $\sum_{i=1}^{N}|w_{i,t} - w_{i,min-c,t}| \leq \delta$ on the respective optimization problems. I consider three threshold parameters: $\delta_1 = 2.5\%$, $\delta_2 = 5\%$ and $\delta_3 = 10\%$.

## 3.3 Performance evaluation and testing

For my main results I use a rolling window estimation approach with window of length $M = 240$ months and a holding period of one month. I choose an estimation window of twenty years to avoid small sample issues for the sample VCV-matrix estimate. Moreover, I assume that at every investment date *t* investors with a risk aversion coefficient of $\gamma = 1$ can only invest in the risky assets and not the factor portfolios or the risk-free asset. Starting from $t = M$, in each month *t* I determine the portfolio weights for the next month for the various diversification strategies by using the most recent *M* observations up to month *t*. For instance, let $w_{k,t}$ be the optimal portfolio weights for some strategy *k*. It follows that the realized excess return over the risk-free rate in month *t+1* is given by

$$R_{k,t+1} = w'_{k,t} R_{t+1} \tag{43}$$

Finally, using $R_{k,t+1}$ I compute the average values of the $T - M$ realized returns $\widehat{\mu_k}$ and standard deviations $\widehat{\sigma_k}$. To study the performance of the various asset allocation models across the datasets summarized in Table 2, I compute three widely-used metrics.

One, the out-of-sample *Sharpe ratio (SR)* of strategy *k*, which is a measure for risk-adjusted return defined by

$$\widehat{SR}_k = \frac{\widehat{\mu_k}}{\widehat{\sigma_k}} \tag{44}$$

To test if the SR of strategy *i* is statistically not different from strategy *j* $H_0: \frac{\widehat{\mu_i}}{\widehat{\sigma_i}} - \frac{\widehat{\mu_j}}{\widehat{\sigma_j}} = 0$, I follow the approach suggested by DeMiguel *et al.* (2009b, p.1928-1929).[4]

Two, I compute the out-of-sample *certainty equivalent (CEQ) return* of strategy *k*, which is given by

$$\widehat{CEQ}_k = \widehat{\mu_k} - \frac{\gamma}{2}\widehat{\sigma_k^2} \qquad (45)$$

with a coefficient of risk aversion $\gamma$. It represents the risk-free rate that an investor is willing to accept in place of investing in a portfolio of risky assets. The difference in CEQ returns of two different strategies *i* and *j* is tested using the delta method following Greene (2012, p.1124) and DeMiguel *et al.* (2009b, p.1929).[5]

Consistent with Tu and Zhou (2011, p.15) and DeMiguel *et al.* (2009b, p.1928) I also report the SR and CEQ performance for an in-sample MV-optimal portfolio, where $M = T$. This is because, even though this strategy is purely hypothetical and not implementable in practice, it serves as a useful benchmark for out-of-sample forecasting and estimation error.

Three, for each strategy the portfolio *turnover*, defined as the average sum of the absolute weight differences across all *N* assets, is computed by

$$turnover_k = \frac{1}{T-M} \sum_{t=1}^{T-M} \sum_{i=1}^{N} |\widehat{w}_{i,k,t+1} - \widehat{w}_{i,k,t^+}| \qquad (46)$$

---

[4] The test statistic $\hat{z}_{SR}$ is computed by

$$\hat{z}_{SR} = \frac{\widehat{\sigma_j}\widehat{\mu_i} - \widehat{\sigma_i}\widehat{\mu_j}}{\sqrt{\hat{\vartheta}}} \xrightarrow{a} N(0,1), \quad where \qquad (47)$$

$$\hat{\vartheta} = \frac{1}{T-M}(2\widehat{\sigma_i^2}\widehat{\sigma_j^2} - 2\widehat{\sigma_i}\widehat{\sigma_j}\widehat{\sigma_{i,j}} + \frac{1}{2}\widehat{\mu_i^2}\widehat{\sigma_j^2} + \frac{1}{2}\widehat{\mu_j^2}\widehat{\sigma_i^2} - \frac{\widehat{\mu_i}\widehat{\mu_j}}{\widehat{\sigma_i}\widehat{\sigma_j}}\widehat{\sigma_{i,j}^2}) \qquad (48)$$

Note that the test statistic $\hat{z}_{SR}$ is only asymptotically standard normal distributed under the assumption that returns are *iid* over time with a normal distribution.

[5] The difference in certainty equivalent returns of two strategies *i* and *j* is given by

$$f(v) = (\mu_i - \frac{\gamma}{2}\sigma_i^2) - (\mu_j - \frac{\gamma}{2}\sigma_j^2) \text{ with } v = (\mu_i \quad \mu_j \quad \sigma_i^2 \quad \sigma_j^2) \qquad (49)$$

and its asymptotic distribution is

$$\sqrt{T}(f(\hat{v}) - f(v)) \xrightarrow{a} N\left(0, \frac{\partial f'}{\partial v} \Theta \frac{\partial f}{\partial v}\right), \quad where \qquad (50)$$

$$\Theta = \begin{pmatrix} \sigma_i^2 & \sigma_{i,j} & 0 & 0 \\ \sigma_{i,j} & \sigma_j^2 & 0 & 0 \\ 0 & 0 & 2\sigma_i^4 & 2\sigma_{i,j}^2 \\ 0 & 0 & 2\sigma_{i,j}^2 & 2\sigma_j^4 \end{pmatrix} \qquad (51)$$

where $\widehat{w}_{i,k,t^+}$ is the portfolio weight *before* rebalancing at $t+1$ and $\widehat{w}_{i,k,t+1}$ is the desired portfolio weight at $t+1$ a*fter* rebalancing. It is important to distinguish here, as for example in the case of the 1/N strategy $w_{i,k,t} = w_{i,k,t+1} = 1/N$, but $w_{i,k,t^+}$ may differ due to changes in asset prices between $t$ and $t+1$ and thus, generally $w_{i,k,t} \neq w_{i,k,t^+}$. This metric can be interpreted as the average percentage of wealth traded in each period. I report absolute turnover for the naïve strategy and for all other strategies relative turnover to that benchmark.

Additionally, *return-loss* per month relative to the 1/N (or minimum-variance) strategy in the presence of proportional transactions costs $c$ is reported. The measure is used to compare the additional return needed for strategy *k* to perform as well as the benchmark in terms of Sharpe ratio. It is computed by

$$return - loss_k = \frac{\tilde{\mu}_{ew}}{\tilde{\sigma}_{ew}} \tilde{\sigma}_k - \tilde{\mu}_k \qquad (52)$$

where $\tilde{\mu}$ and $\tilde{\sigma}$ correspond to the out-of-sample mean and volatility of *net excess returns* $\tilde{R}_{k,t+1}$, respectively. Net excess returns after transaction costs $\tilde{R}_{k,t+1}$ of strategy *k* at time $t+1$ are given by the evolution of wealth $W_{k,t}$ net of transaction costs

$$\tilde{R}_{k,t+1} = \frac{W_{k,t+1}}{W_{k,t}} - 1, \qquad where \qquad (53)$$

$$W_{k,t+1} = W_{k,t}(1 + R_{k,t+1})(1 - c \sum_{i=1}^{N} |\widehat{w}_{i,k,t+1} - \widehat{w}_{i,k,t^+}|) \qquad (54)$$

in which I choose $c$ to be 50 basis points per transaction to be consistent with DeMiguel *et al.* (2009b).

# 4 Empirical results

In this section I discuss the empirical performance of the various allocation rules across all datasets. I compare them to the naïve but also the minimum-variance strategy, as it has become a popular benchmark in the literature. To assess performance, I compute Sharpe ratios (Table 3), CEQ returns (Table 6 in Appendix C) and turnover (Table 7 in Appendix C). In each of the following tables I examine the various strategies (rows) across datasets (columns).

## 4.1 Sharpe ratios

The first row in Table 3 reports the Sharpe ratios for the 1/N benchmark. The second and third row give the performance for the in- and out-of-sample MV-strategy, respectively. By construction, when there is no estimation or forecasting error, the in-sample MV Sharpe ratios are always the highest among all strategies. However, note that the magnitude of the difference between the in-sample and the out-of-sample SR for the MV-portfolio is substantial. For instance, for the dataset 'International', the MV out-of-sample performance ($SR = 0.1375$) is less than half of the in-sample benchmark ($SR = 0.3869$). Note that this difference in performance may arise due to both forecasting error caused by random shocks or moment estimation error. Therefore, a comparison of the optimal MV portfolio and the estimation-free naïve strategy helps to assess the approximate impact of estimation error on performance.

The equally-weighted portfolio achieves out-of-sample ($SR = 0.1563$) only a quarter of the performance of the in-sample MV strategy ($SR = 0.5918$) for the 'FF-4' dataset, for example. However, I find that the out-of-sample performance of the MV-portfolio is generally lower than the estimation-free naïve strategy for all datasets except for the 'International' and 'FF-4' data. Hence, estimation errors seem to eliminate the benefits of optimal diversification in most cases. This seems to be particularly true for individual stock portfolios, since I find that the Sharpe ratio of the sample MV strategy ($SR = 0.0461$) is significantly lower than the 1/N rule ($SR = 0.1985$) for the 'DJIA' dataset, for instance.

Overall, I can confirm the well-known poor performance of the classical mean-variance strategy in an out-of-sample setting. My findings are largely in line with DeMiguel *et al.* (2009b, p.1930-1932), who also conclude that naïve diversification is generally superior to the classical approach.

Moreover, the minimum-variance portfolio has become a popular benchmark in the literature and industry due its relatively good performance and the predictability of second moments. However, I find that the restriction of moments alone in the 'min' strategy cannot

**Table 3 | Sharpe ratios**

This table reports the monthly Sharpe ratios for the various portfolio selection strategies listed in Table 1. The p-value in parentheses refers to the two-tailed test of the difference between the Sharpe ratio of each strategy from that of the naïve benchmark (ew). For the 'DJIA' dataset the number of assets $N$ refers to the average number of securities available at each investment date $t$. I only report results for the best performing parameter choices.

| Strategy | Industry $N=10+1$ | International $N=8+1$ | SMB/HML/UMD $N=3+1$ | FF-4 $N=25+4$ | DJIA $N=33+4$ | SP500 $N=60+4$ |
|---|---|---|---|---|---|---|
| ew | 0.1736 | 0.0964 | -0.0134 | 0.1563 | 0.1985 | 0.1608 |
| mv (in-sample) | 0.2390 | 0.3869 | 0.1027 | 0.5918 | - | - |
| mv | 0.1208 | 0.1375 | -0.0300 | 0.4097 | 0.0461 | 0.0551 |
|  | (0.32) | (0.86) | (0.81) | (0.00) | (0.03) | (0.13) |
| bs | 0.1855 | 0.1571 | -0.0387 | 0.4373 | 0.0858 | 0.1209 |
|  | (0.78) | (0.77) | (0.70) | (0.00) | (0.07) | (0.53) |
| min | 0.2284 | 0.0590 | -0.0204 | 0.3494 | 0.1860 | 0.1993 |
|  | (0.13) | (0.79) | (0.77) | (0.00) | (0.78) | (0.46) |
| mv-var ($\delta = 5\%$) | 0.2089 | 0.0711 | -0.0121 | 0.2032 | 0.1964 | 0.2331 |
|  | (0.22) | (0.72) | (0.95) | (0.02) | (0.95) | (0.05) |
| rrt ($\omega = 2$) | 0.1445 | 0.2341 | -0.0595 | 0.1584 | 0.1956 | 0.2002 |
|  | (0.02) | (0.12) | (0.29) | (0.76) | (0.84) | (0.06) |
| mv-min | 0.2092 | 0.1582 | 0.0351 | 0.4355 | 0.1434 | 0.1973 |
|  | (0.35) | (0.72) | (0.47) | (0.00) | (0.27) | (0.49) |
| ew-min | 0.2289 | 0.0557 | 0.0109 | 0.3391 | 0.1893 | 0.2000 |
|  | (0.11) | (0.70) | (0.73) | (0.00) | (0.83) | (0.45) |
| ew-mv | 0.1476 | 0.1406 | 0.0494 | 0.4125 | 0.1225 | 0.1438 |
|  | (0.46) | (0.62) | (0.35) | (0.00) | (0.13) | (0.51) |
| ew-mv-min | 0.2011 | 0.1149 | 0.0327 | 0.0652 | 0.0669 | 0.2068 |
|  | (0.46) | (0.05) | (0.49) | (0.12) | (0.04) | (0.29) |
| ac-mv | 0.1283 | 0.1582 | 0.0437 | 0.3336 | 0.0640 | 0.0056 |
|  | (0.37) | (0.71) | (0.14) | (0.01) | (0.06) | (0.03) |
| ac-bs | 0.1907 | 0.1102 | 0.0255 | 0.3445 | 0.1054 | 0.0630 |
|  | (0.67) | (0.92) | (0.25) | (0.00) | (0.15) | (0.15) |
| ac-mv-min | 0.2109 | 0.0862 | 0.0264 | 0.3402 | 0.1591 | 0.1865 |
|  | (0.31) | (0.94) | (0.25) | (0.00) | (0.44) | (0.63) |
| mv-c | 0.1387 | 0.1454 | 0.0382 | 0.1628 | 0.1639 | 0.0815 |
|  | (0.20) | (0.56) | (0.12) | (0.72) | (0.38) | (0.02) |
| min-c | 0.2044 | 0.0698 | -0.0204 | 0.1950 | 0.1949 | 0.2235 |
|  | (0.28) | (0.69) | (0.77) | (0.06) | (0.91) | (0.08) |
| bs-c | 0.1785 | 0.1368 | 0.0320 | 0.1714 | 0.1879 | 0.1097 |
|  | (0.84) | (0.53) | (0.14) | (0.38) | (0.76) | (0.11) |
| ac-mv-c | 0.1741 | 0.1390 | 0.0393 | 0.178 | 0.2045 | 0.1107 |
|  | (0.99) | (0.51) | (0.06) | (0.15) | (0.85) | (0.11) |
| ac-bs-c | 0.1901 | 0.0906 | 0.0275 | 0.1816 | 0.2067 | 0.1781 |
|  | (0.50) | (0.92) | (0.10) | (0.10) | (0.79) | (0.60) |
| ac-mv-min-c | 0.1971 | 0.0653 | 0.0282 | 0.1819 | 0.1985 | 0.2245 |
|  | (0.37) | (0.63) | (0.10) | (0.12) | (1.00) | (0.08) |
| min-norm ($\delta = 10\%$) | 0.2085 | 0.0759 | -0.0203 | 0.2052 | 0.1932 | 0.2298 |
|  | (0.24) | (0.78) | (0.77) | (0.03) | (0.88) | (0.07) |
| bs-norm ($\delta = 10\%$) | 0.2020 | 0.0587 | -0.0094 | 0.2034 | 0.1967 | 0.2271 |
|  | (0.33) | (0.62) | (0.83) | (0.03) | (0.96) | (0.11) |

outperform the naïve strategy in half of the datasets, where the performance improvement is only significant for the 'FF-4' data.[6] Hence, the empirical evidence in favour of the 'min'-portfolio is less convincing than results by DeMiguel *et al.* (2009b, p.1932-p.1933), who find superior performance in most of their datasets. Due to this empirical observation I only report p-values of the difference to the 1/N benchmark.

Additionally, it is important to note that the 'bs'-strategy is generally inferior to the naïve and minimum-variance portfolios in most datasets. However, it achieves larger Sharpe ratios than the sample MV-portfolio in all datasets except for 'SMB/HML/UMD'. As the Sharpe ratios are considerably larger in most scenarios, my results clearly suggest that Bayesian strategies can be effective in dealing with estimation errors and are preferable over the sample MV-portfolio.

However, the weak performance of optimal portfolios relative to the naïve strategy is a well-known observation, which DeMiguel *et al.* (2009b) also document for a range of other sophisticated asset allocation strategies that were originally designed to reduce estimation error. Hence, the interesting question of this dissertation is if sophisticated diversification models proposed in the past decade have particularly improved in terms of reducing estimation error of expected returns.

In general, the SR performance ranking of the optimal strategies across datasets as depicted in Figure 1 in Appendix C reveals two patterns. First, none of the recently proposed strategies can outperform the 1/N rule or the minimum-variance portfolio consistently. I find that 1/N is outperformed by sophisticated models only in 3.7 datasets on average while this number is 2.9 datasets for the minimum-variance benchmark. Second, while some strategies rank consistently low and are generally dominated by the naïve and minimum-variance strategy, others show superior performance in at least four or five of the datasets considered.

Broadly in line with results by DeMiguel *et al.* (2014) I find that exploiting the serial dependence of monthly stock returns in the 'mv-var'-strategy results in performance improvements. It achieves higher Sharpe ratios than the 'ew', 'mv' and 'min' portfolios in four datasets and the difference to the naïve benchmark is significant for the 'FF-4' and 'SP500' data. This contrasts with the literature which finds that it is optimal to ignore expected returns and extends findings by DeMiguel *et al.* to monthly data as they only focus on modelling daily returns. But, this result is not surprising given that I also find significant serial correlation at various lags for many risky assets in the asset universes, which is apparently exploited by the VAR-model (see Table 5 in Appendix B). However, as fewer risky assets in the 'International' and 'DJIA' datasets exhibit significant serial dependence, the 'mv-var' Sharpe ratios of $SR = 0.0711$ and $SR = 0.1964$ respectively are lower than the ones for the naïve portfolio ($SR = 0.0964$ and $SR = 0.1985$, respectively).

---

[6] Note that the weak performance of the minimum-variance portfolio is likely due to the fact that I am only using a sample estimate of the VCV-matrix, while better estimators have already been proposed in the literature by Ledoit and Wolf (2003; 2017), for instance.

Moreover, I find no support for the reward-to-risk timing strategy suggested by Kirby and Ostdiek (2012), as it performs worse than the minimum-variance approach in most datasets. Further, the Sharpe ratio of the 'rrt'-rule is significantly lower ($SR = 0.1445$) than the naïve strategy ($SR = 0.1736$) for the 'Industry' dataset. Since I do not observe better performance for individual stocks that exhibit higher variation in returns than the portfolio datasets (see Table 4 in Appendix B), my results contrast with the explanation by Kirby and Ostdiek (2012, p.456). They argue that the 'rrt' strategy should outperform 1/N in cases where the cross-sectional variation of expected returns is large, which is not confirmed.

Among the unconstrained combination rules I find that only the 'three-fund' model by Kan and Zhou (2007) and the combination of the naïve and minimum-variance strategy by DeMiguel *et al.* (2009b) achieve better performance over the 1/N or minimum-variance benchmark in most datasets. The 'ew-min'-strategy outperforms the 'ew', 'mv' and 'min' portfolios in four datasets. For instance, the difference to the naïve benchmark is significant for the 'FF-4' data, where it has a Sharpe ratio ($SR = 0.3391$) twice as large as the one for the 1/N rule ($SR = 0.1563$). However, for the same dataset the 'mv-min'-model has an even higher and significant Sharpe ratio of $SR = 0.4355$ compared to the 1/N rule. Additionally, it outperforms the naïve strategy in all datasets except for the 'DJIA' data, where its performance is statistically indistinguishable from the benchmark.

Both the optimal combination strategies by Tu and Zhou (2011) and the unconstrained robust portfolios by Anderson and Cheng (2016) are inferior to the minimum-variance strategy in most datasets. Anderson and Cheng's robust portfolios can only improve the non-robust versions in half of the cases, which is a weaker result compared to their empirical findings (Anderson and Cheng, 2016, p.1353).[7] Furthermore, even though the 'ew-mv' strategy by Tu and Zhou (2011) has strictly higher Sharpe ratios than the classical 'mv' rule, it fails to dominate the 'ew' and 'min'-strategy. Moreover, the 'ew-mv-min' model has lower Sharpe ratios than its 'mv-min' counterpart in all but one dataset. These findings are in sharp contrast to results documented by Tu and Zhou (2011, p.28), who report that it performed best among all the models they considered.

Like DeMiguel *et al.* (2009b, p.1933) I find that imposing short-sale constraints alone does not necessarily improve performance. Except for the short-sale constrained mean-variance portfolio, the 'min-c' and 'bs-c' strategies have lower Sharpe ratios in most scenarios than their unconstrained counterparts. However, when imposing short-sale constraints on the robust portfolios by Anderson and Cheng (2016), I find that the 'ac-mv-c' and 'ac-bs-c' strategies achieve better performance than the 1/N rule in all but one dataset. Moreover, the 'ac-mv-min-c' model outperforms the 'ew', 'mv' and 'min' portfolios in four datasets. For instance, the difference to the naïve benchmark is significant for the 'SP500' data, where it has a Sharpe ratio of $SR = 0.2245$ compared to $SR = 0.1608$ for the 1/N rule.

---

[7] However, they find further performance gains by combining the robust portfolios using a Bayesian averaging approach.

Additionally, I find that both the minimum-variance and Bayes-Stein portfolios with norm constraints achieve higher Sharpe ratios in most datasets than the unconstrained and short-sale constrained versions. For instance, the 'min-norm' strategy shows superior performance compared to the 'min-c' rule in all datasets except the 'DJIA' data and also outperforms the 'min' strategy in four of them. This also suggests that the good performance of the 'mv-var' strategy is partly due to the norm-constraints on portfolio weights.

Finally, I find no empirical support for the conclusions of DeMiguel *et al.* (2009b, p.1920), who argue based on their simulation experiments that the 1/N strategy is likely to outperform optimal diversification schemes when the number of risky assets *N* is large. This is because when comparing the last two columns in Table 3, twelve sophisticated models outperform the naïve benchmark (of which seven exhibit statistically significant difference) in the 'SP500' dataset while it is only two (though insignificant) for the 'DJIA' portfolios.[8] Moreover, I cannot empirically support their general claim that 1/N tends to underperform when levels of idiosyncratic volatility are high, as on average fewer optimal strategies are able to outperform naïve diversification with stock portfolios.

## 4.2 Certainty equivalent returns

The comparison of CEQ returns in Table 6 in Appendix C strengthens the general conclusions from the analysis of Sharpe ratios. The CEQ returns of the in-sample MV-portfolio are the largest among all strategies and none of the optimal asset allocation schemes can outperform the 'ew' or 'min'-portfolios consistently. In fact, as depicted in Figure 2 in Appendix C most models deteriorate in performance relative to the 1/N benchmark as they outperform it less often (only in 3.3 datasets on average). Significant superior performance is only found in the 'International' and 'FF-4' dataset. However, the 'mv-var'-strategy remains strong and achieves higher CEQ returns than the 'ew', 'mv' and 'min' portfolios in four datasets (although these are insignificant). Further, like in the previous section the 'mv-min' model outperforms 1/N in all datasets except 'DJIA'. For instance, it achieves a significantly better performance ($CEQ = 0.0188$) relative to the naïve portfolio ($CEQ = 0.0064$) for the 'FF-4' dataset.

## 4.3 Portfolio turnover

In Table 7 in Appendix C I report portfolio turnover for all strategies, where the first line reports absolute turnover for the naïve portfolio and turnover relative to the 1/N rule for all other strategies. Comparing absolute turnover of the 1/N rule across datasets, the relatively high average wealth per month traded of about 68% for the 'SP500' dataset stands out. This however, is not surprising as the sixty assets are randomly sampled at every investment date *t* and therefore, securities are often completely sold off and only hold for a single period.

---
[8] The 'SP500" portfolios consist on average of twice as many assets as the 'DJIA' portfolios.

From Figure 3 in Appendix C, it is clear that portfolio turnover of the optimal strategies is generally much higher than for the 1/N benchmark. However, it is also easy to see that turnover for the 'rrt' and constrained optimal strategies is in many scenarios only slightly higher than the naïve benchmark.

Finally, turnover for the 'mv', 'bs' and unconstrained combination strategies is generally substantially larger than for the 1/N rule, whereas it is only moderately more for the 'mv-var' and 'min' strategy.

In Table 8 in Appendix C I report return-loss relative to the 1/N strategy, based on returns net of proportional transaction costs as defined in Eq. 52. When comparing these to the Sharpe ratios in Table 3, note that all strategies with negative return-loss also by construction have higher Sharpe ratios than the 'ew'-portfolio. A negative return-loss implies that even in the presence of transaction costs these strategies attain a higher SR than the naïve benchmark. In fact, only in three cases, where an optimal strategy yields a higher Sharpe ratio than 1/N, high turnover and transaction costs lead to an actual return-loss relative to 1/N. I find that in most scenarios the optimal strategies have negative return-loss (*i.e.* in about 58% of the scenarios).

However, this picture changes drastically when considering return-loss relative to the minimum-variance strategy, where most optimal diversification rules exhibit an actual return-loss (see Table 9 in Appendix C).

### 4.4 Limitations

It is important to note that my statistical testing procedures assume that returns are distributed independently and identically (*iid*) over time. However, this assumption is generally violated in my empirical datasets, as according to Ljung-Box test results I find significant serial correlation at the 1% significance level for many assets (see Table 5 in Appendix B). Hence, without robust testing procedures any of my significant results need to be treated with wariness. This however, is not a big issue since even the non-robust tests fail to be significant consistently.

Besides the significance of test results, other limitations include the weak sample estimate of second moments used in this thesis. This involves the consideration of small sample issues, which may occur if the number of observations *M* is less than one order of magnitude of the number of assets *N* (Ledoit and Wolf, 2003, p.604). This applies particularly to my results for the individual stock data, where $M = 240$ observations are likely too few for the relatively large number of assets. Moreover, considering the short-run dynamics of stock volatility, my estimate of second moments using twenty years of monthly returns is likely a weak predictor of next period risk. Instead, future research may consider using a more recent and shorter window of daily data for estimating the variance-covariance of returns, for example.

# 5   Summary and conclusions

In the past decade many researchers have tried to overcome the limitations of the classical mean-variance optimization strategy. This is because past research shows that the Markowitz and other sophisticated optimization rules cannot consistently outperform the simple equally-weighted portfolio in out-of-sample benchmarks. Hence, I test empirically if researchers have recently made progress in developing new optimal diversification rules that are able to beat not only the naïve but also the minimum-variance strategy.

To evaluate out-of-sample performance of the sixteen strategies considered across six empirical datasets, I compute optimal portfolio weights in a rolling-window procedure and report out-of-sample Sharpe ratios, certainty-equivalent returns and portfolio turnover. Moreover, I test whether the performance of the optimal strategies is significantly different from the 1/N benchmark.

As measured by the Sharpe ratio or certainty equivalent return most recently proposed optimal diversification schemes outperform the naïve portfolio, but generally not so in a consistent or statistically significant way on conventional significance levels. Moreover, they are generally not able to achieve higher risk-adjusted returns than the minimum-variance approach, and are less attractive due to high portfolio turnover.

While these results are in line with DeMiguel *et al.* (2009b), they contrast with findings by researchers who originally proposed the tested models and, therefore, weaken their claim of superior performance (Anderson and Cheng, 2016; DeMiguel *et al.*, 2014; Kirby and Ostdiek, 2012; Tu and Zhou, 2011). It seems that errors in estimating particularly expected returns are still eroding the benefits of portfolio optimization, and simpler diversification rules are not inefficient after all.

My results have two important implications. First, optimal portfolio selection remains a daunting task due to the difficulty of estimating expected returns particularly. Therefore, further research effort should be devoted to both improving the estimation of expected returns and exploring diversification rules that do not require the estimation of expected returns directly, but also use other available information about the stock's characteristics. Second, based on my findings, incorporating weight constraints into combination rules such as the 'three-fund' model by Kan and Zhou (2007) may be a promising research direction to pursue. This is because weight constraints are likely to reduce turnover and may also improve performance.

# Appendix

## A  Description of empirical datasets

This appendix describes the six empirical datasets considered in this study. Each dataset contains monthly excess returns over the one-month US T-Bill return from the Kenneth French Data Library.[9]

### A.1  Industry portfolios

The 'Industry' dataset consists of ten value-weighted industry portfolios in the United States. The industries are consumer non-durables, consumer durables, manufacturing, energy, high-tech, telecommunications, retail, healthcare, utilities and others. The data ranges from 06/1963 to 03/2018 and was retrieved from the Kenneth French Data Library. The dataset is being augmented using the excess returns on the US equity market portfolio, MKT.

### A.2  International portfolios

The 'International' dataset consists of eight international MSCI equity indices denoted in their respective local currencies: Canada, France, Germany, Italy, Japan, Switzerland, the UK and the US. Moreover, the US-dollar MSCI World index is used as the market factor portfolio. Returns are computed based on the month-end value of the equity index for the period from 02/1995 to 03/2018. The MSCI data was retrieved from Bloomberg.

### A.3  MKT, SMB, HML and UMD portfolios

The factors dataset contains monthly factor returns from 06/1963 to 03/2018 on the US equity market portfolio (MKT) and the SMB, HML and Momentum (UMD) zero-cost portfolios (Fama and French, 1993; Carhart, 1997). Additionally, it contains the one-month US T-Bill return (RF) used as the risk-free asset in this thesis. The data is taken from Kenneth French Data Library.

### A.4  25 Size- and book-to-market-sorted portfolios

The data consists of 25 portfolios sorted by size and book-to-market ratio. The monthly value-weighted returns ranging from 06/1963 to 03/2018 are taken from Kenneth French Data Library. I augment the sorted portfolios dataset with the four factor portfolios: MKT, HML, SMB and UMD, which are taken from Kenneth French Data Library as well.

### A.5  Individual stock data

I retrieve S&P 500 and Dow Jones Industrial Average (DJIA) index constituents data from Compustat/CRSP database. Based on the stock-level CUSIP identifier provided by Standard

---

[9] Kenneth French data retrieved in May 2018 from
http://mba.tuck.dartmouth.edu/pages/faculty/ken.french/data_library.html.

& Poor's I first retrieve all historical index constituents in the period from 06/1963 – 03/2018. Next, I obtain the month-end holding period returns including dividends for all constituents. To avoid survivorship bias, all constituents which have historically been in the index for at least one day and have at least 241 months of returns data are included. Further, due to changes in the index composition the available securities may vary at each investment date *t* and some stocks do not contain observations for the full time-period. Finally, I augment each individual stock dataset with the four factor portfolios: MKT, HML, SMB and UMD, which are taken from Kenneth French Data Library.

# B  Data summary

## B.1  Summary statistics

**Table 4 | Summary statistics**

This table lists properties of the monthly returns data used in this paper. The second column gives short labels for the respective asset universes from which portfolios are selected. The third and fourth column provide the total number of risky assets $N$ in the respective universe of securities and the total number of monthly observations $T$. Please note that the sample size $T$ may vary for some risky assets $i$ in the asset universes 5 and 6. Columns 5 and 6 show the beginning and the end of the data. Columns 7 through 16 calculate some key statistics and show its lowest and highest values over all risky assets in the respective universe. Except for kurtosis and skewness all statistics are report in percentage. In this table the risk-free rate is not subtracted from the nominal returns.

| # | Dataset | $N$ | $T$ | Start date | End date | Lowest mean (%) | Highest mean (%) | Lowest std (%) | Highest std (%) | Lowest return (%) | Highest return (%) | Lowest kurtosis | Highest kurtosis | Lowest skewness | Highest skewness |
|---|---|---|---|---|---|---|---|---|---|---|---|---|---|---|---|
| 1 | Industry | 10 | 658 | 1963-06 | 2018-03 | 0.821 | 1.068 | 3.987 | 6.370 | -32.630 | 42.630 | 1.106 | 4.883 | -0.477 | 0.132 |
| 2 | International | 8+1 | 278 | 1995-02 | 2018-03 | 0.191 | 0.718 | 3.831 | 6.106 | -24.926 | 21.792 | 0.448 | 2.797 | -0.805 | 0.137 |
| 3 | 4-Factors | 4 | 658 | 1963-06 | 2018-03 | 0.216 | 0.667 | 2.811 | 4.385 | -34.390 | 22.140 | 1.994 | 10.654 | -1.340 | 0.513 |
| 4 | Sorted 25 | 25 | 658 | 1963-06 | 2018-03 | 0.655 | 1.480 | 4.231 | 7.830 | -34.216 | 41.052 | 1.339 | 3.696 | -0.559 | 0.044 |
| 5 | SP500 | 938 | 658 | 1963-06 | 2018-03 | -0.265 | 4.186 | 4.776 | 27.976 | -98.130 | 380.000 | -0.042 | 117.234 | -3.203 | 8.471 |
| 6 | DJIA | 42 | 658 | 1963-06 | 2018-03 | 0.330 | 2.535 | 4.920 | 15.720 | -83.481 | 244.977 | 0.374 | 117.234 | -0.292 | 7.642 |

## B.2 Autocorrelation tests

**Table 5 | Autocorrelation tests**

This table reports Ljung-Box test results for the null hypothesis of no autocorrelation. The second column gives short labels for the respective asset universes. The third and fourth column provide the total number of risky assets *N* in the respective universe of securities and the total number of monthly observations *T*. The last four columns report the number of securities in the respective asset universe with significant rejection of the null at a significance level of $a = 1\%$ for different lag lengths.

| # | Dataset | *N* | *T* | Number of lags | | | |
|---|---------|-----|-----|----|----|----|----|
|   |         |     |     | 2  | 4  | 8  | 16 |
| 1 | Industry | 10 | 658 | 3 | 1 | 2 | 2 |
| 2 | International | 8+1 | 278 | 2 | 0 | 0 | 0 |
| 3 | 4-Factors | 4 | 658 | 1 | 1 | 1 | 0 |
| 4 | Sorted 25 | 25 | 658 | 15 | 9 | 8 | 9 |
| 5 | SP500 | 938 | 658 | 81 | 83 | 115 | 135 |
| 6 | DJIA | 42 | 658 | 3 | 4 | 6 | 9 |

# C  Further results

## C.1  Sharpe ratios ranking

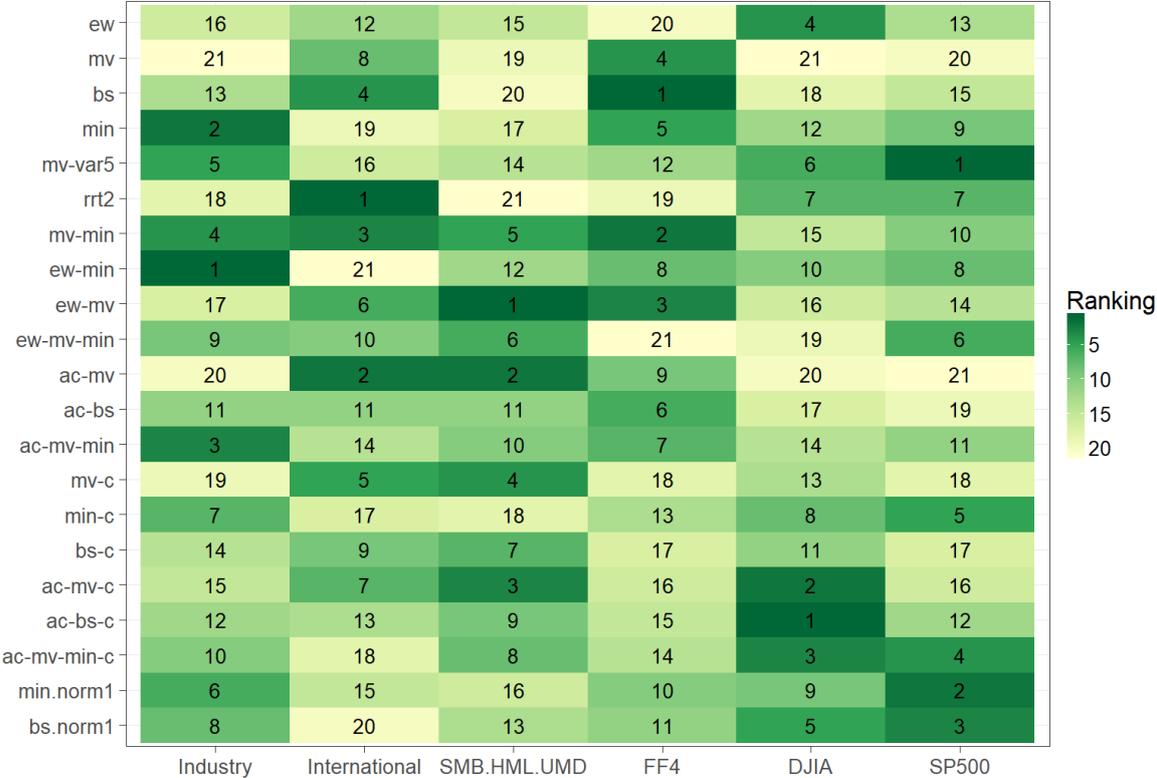

**Figure 1 | Sharpe ratio ranking**

This figure depicts the monthly Sharpe ratio rankings for the various portfolio selection strategies. For each dataset (x-axis) the Sharpe ratios of the portfolio rules (y-axis) are ranked from best (*dark green*) to worst (*bright yellow*) performance.

## C.2  Certainty-equivalent returns

**Table 6 | Certainty-equivalent returns**

This table reports the monthly CEQ return for the various portfolio selection strategies listed in Table 1. The p-value in parentheses refers to the two-tailed test of the difference between the CEQ return of each strategy from that of the naïve benchmark (ew).

| Strategy | Industry<br>N=10+1 | International<br>N=8+1 | SMB/HML/UMD<br>N=3+1 | FF-4<br>N=25+4 | DJIA<br>N=33+4 | SP500<br>N=60+4 |
|---|---|---|---|---|---|---|
| ew | 0.0062 | 0.0026 | -0.0004 | 0.0064 | 0.0082 | 0.0065 |
| mv (in-sample) | 0.0081 | 0.0313 | 0.0045 | 0.0296 | - | - |
| mv | 0.0057 | 0.0094 | -0.0052 | 0.0259 | -0.0018 | 0.0002 |
| | (0.82) | (0.38) | (0.12) | (0.00) | (0.05) | (0.17) |
| bs | 0.0070 | 0.0064 | -0.0025 | 0.0214 | 0.0035 | 0.0054 |
| | (0.58) | (0.21) | (0.23) | (0.00) | (0.11) | (0.68) |
| min | 0.0074 | 0.0012 | -0.0005 | 0.0119 | 0.0066 | 0.0068 |
| | (0.26) | (0.36) | (0.86) | (0.00) | (0.30) | (0.86) |
| mv-var ($\delta=5\%$) | 0.0067 | 0.0015 | -0.0003 | 0.0075 | 0.0067 | 0.0071 |
| | (0.59) | (0.20) | (0.81) | (0.16) | (0.19) | (0.62) |
| rrt ($\omega=2$) | 0.0056 | 0.0064 | -0.0017 | 0.0066 | 0.0083 | 0.0077 |
| | (0.19) | (0.00) | (0.08) | (0.39) | (0.74) | (0.11) |
| mv-min | 0.0072 | 0.0043 | 0.0006 | 0.0188 | 0.0053 | 0.0068 |
| | (0.40) | (0.37) | (0.59) | (0.00) | (0.10) | (0.87) |
| ew-min | 0.0074 | 0.0011 | -0.0135 | 0.0114 | 0.0067 | 0.0068 |
| | (0.25) | (0.21) | (0.06) | (0.00) | (0.30) | (0.85) |
| ew-mv | 0.0056 | 0.0040 | -0.2713 | 0.0225 | 0.0051 | 0.0055 |
| | (0.58) | (0.17) | (0.00) | (0.00) | (0.13) | (0.29) |
| ew-mv-min | 0.0071 | 0.0031 | 0.0004 | 0.0019 | -0.0327 | 0.0069 |
| | (0.47) | (0.00) | (0.39) | (0.19) | (0.00) | (0.81) |
| ac-mv | 0.0055 | 0.0039 | 0.001 | 0.0336 | 0.0011 | -0.0092 |
| | (0.70) | (0.47) | (0.29) | (0.00) | (0.13) | (0.01) |
| ac-bs | 0.0068 | 0.0022 | 0.0003 | 0.0284 | 0.0046 | 0.0019 |
| | (0.64) | (0.82) | (0.37) | (0.00) | (0.22) | (0.17) |
| ac-mv-min | 0.0071 | 0.0017 | 0.0003 | 0.0251 | 0.0057 | 0.0064 |
| | (0.43) | (0.57) | (0.40) | (0.00) | (0.17) | (0.97) |
| mv-c | 0.0049 | 0.0032 | 0.0007 | 0.0069 | 0.0083 | 0.0030 |
| | (0.14) | (0.51) | (0.34) | (0.46) | (0.92) | (0.02) |
| min-c | 0.0066 | 0.0015 | -0.0005 | 0.0072 | 0.0067 | 0.0068 |
| | (0.70) | (0.17) | (0.86) | (0.30) | (0.17) | (0.80) |
| bs-c | 0.0060 | 0.0030 | 0.0005 | 0.0072 | 0.0079 | 0.0039 |
| | (0.74) | (0.58) | (0.34) | (0.21) | (0.86) | (0.02) |
| ac-mv-c | 0.0059 | 0.0031 | 0.0007 | 0.0072 | 0.0083 | 0.0038 |
| | (0.61) | (0.54) | (0.13) | (0.14) | (0.92) | (0.02) |
| ac-bs-c | 0.0062 | 0.0019 | 0.0004 | 0.0071 | 0.0073 | 0.0057 |
| | (0.93) | (0.37) | (0.14) | (0.21) | (0.45) | (0.48) |
| ac-mv-min-c | 0.0063 | 0.0013 | 0.0004 | 0.0070 | 0.0067 | 0.0068 |
| | (0.90) | (0.11) | (0.15) | (0.31) | (0.19) | (0.80) |
| min-norm ($\delta=10\%$) | 0.0067 | 0.0016 | -0.0005 | 0.0075 | 0.0066 | 0.0070 |
| | (0.62) | (0.28) | (0.86) | (0.17) | (0.17) | (0.72) |
| bs-norm ($\delta=10\%$) | 0.0064 | 0.0012 | -0.0003 | 0.0075 | 0.0067 | 0.0069 |
| | (0.82) | (0.11) | (0.68) | (0.18) | (0.20) | (0.76) |

## C.3 Certainty-equivalent returns ranking

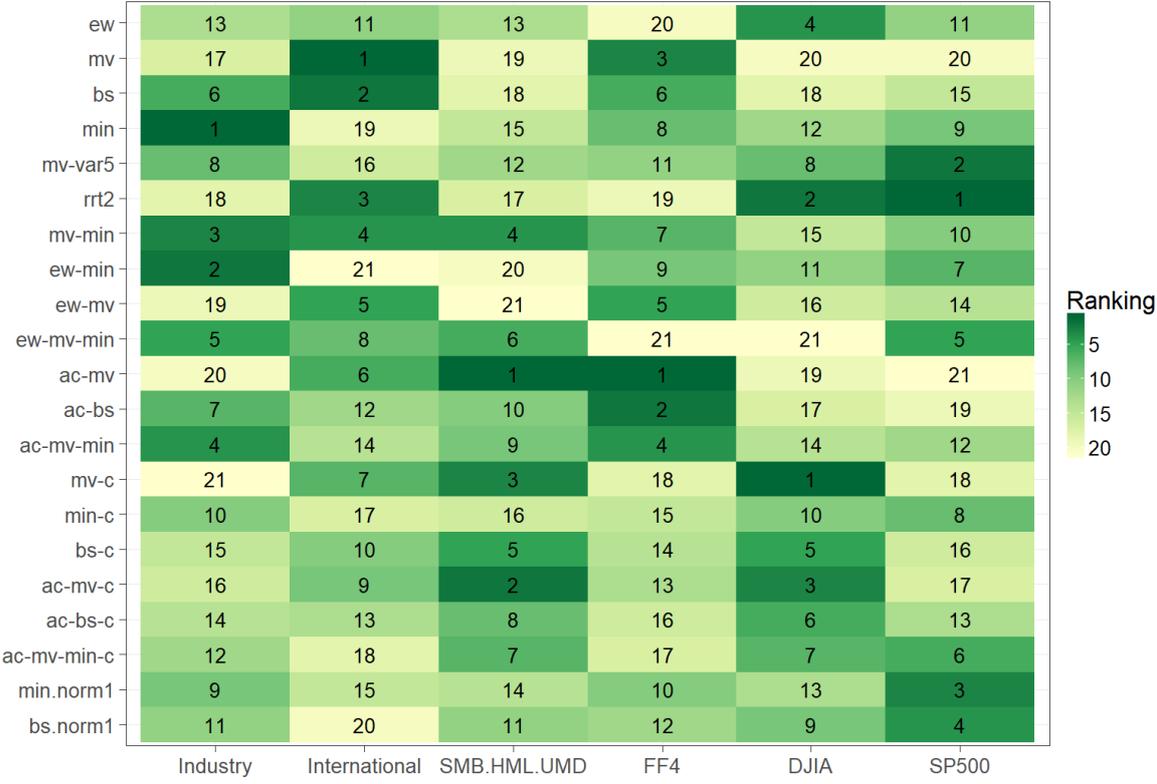

**Figure 2 | Certainty-equivalent return ranking**

This figure depicts the monthly certainty-equivalent (CEQ) return rankings for the various portfolio selection strategies. For each dataset (x-axis) the CEQ returns of the portfolio rules (y-axis) are ranked from best (*dark green*) to worst (*bright yellow*) performance.

## C.4 Turnover

**Table 7 | Relative monthly turnover**

For each empirical dataset the table reports absolute monthly turnover for the naïve strategy in the first line. For all remaining strategies, it reports turnover relative to the 1/N benchmark.

| Strategy | Industry $N=10+1$ | International $N=8+1$ | SMB/HML/UMD $N=3+1$ | FF-4 $N=25+4$ | DJIA $N=33+4$ | SP500 $N=60+4$ |
|---|---|---|---|---|---|---|
| ew | 0.0256 | 0.0424 | 0.0234 | 0.0194 | 0.0513 | 0.6831 |
| mv | 25.24 | 104.79 | 32.43 | 107.13 | 29.17 | 10.01 |
| bs | 10.06 | 36.96 | 9.13 | 67.91 | 12.20 | 5.40 |
| min | 3.73 | 2.64 | 0.93 | 21.64 | 3.32 | 3.54 |
| mv-var ($\delta=5\%$) | 3.81 | 2.42 | 1.33 | 5.49 | 2.80 | 1.33 |
| rrt ($\omega=2$) | 1.03 | 0.82 | 0.78 | 3.01 | 2.18 | 1.43 |
| mv-min | 6.38 | 21.06 | 97.05 | 50.33 | 5.30 | 3.55 |
| ew-min | 3.54 | 2.74 | 110.27 | 19.09 | 3.22 | 3.52 |
| ew-mv | 11.85 | 11.59 | 324.57 | 76.68 | 7.74 | 2.27 |
| ew-mv-min | 12.57 | 3.61 | 26.83 | 523.73 | 132.18 | 2.94 |
| ac-mv | 18.38 | 9.90 | 3.15 | 200.15 | 21.47 | 17.03 |
| ac-bs | 7.95 | 4.74 | 1.95 | 131.47 | 9.17 | 7.21 |
| ac-mv-min | 5.53 | 3.53 | 2.16 | 105.06 | 4.18 | 3.59 |
| mv-c | 5.83 | 3.63 | 1.24 | 9.76 | 2.09 | 1.40 |
| min-c | 1.34 | 1.14 | 0.93 | 2.13 | 1.13 | 1.20 |
| bs-c | 3.56 | 2.31 | 1.62 | 8.35 | 2.31 | 1.31 |
| ac-mv-c | 3.68 | 2.28 | 1.29 | 6.65 | 2.06 | 1.30 |
| ac-bs-c | 2.19 | 2.36 | 1.18 | 4.99 | 1.49 | 1.23 |
| ac-mv-min-c | 1.64 | 1.92 | 1.24 | 4.31 | 1.28 | 1.20 |
| min-norm ($\delta=10\%$) | 2.78 | 2.31 | 0.93 | 4.33 | 1.62 | 1.35 |
| bs-norm ($\delta=10\%$) | 2.81 | 2.59 | 0.95 | 5.30 | 1.73 | 1.36 |

## C.5 Absolute turnover ranking

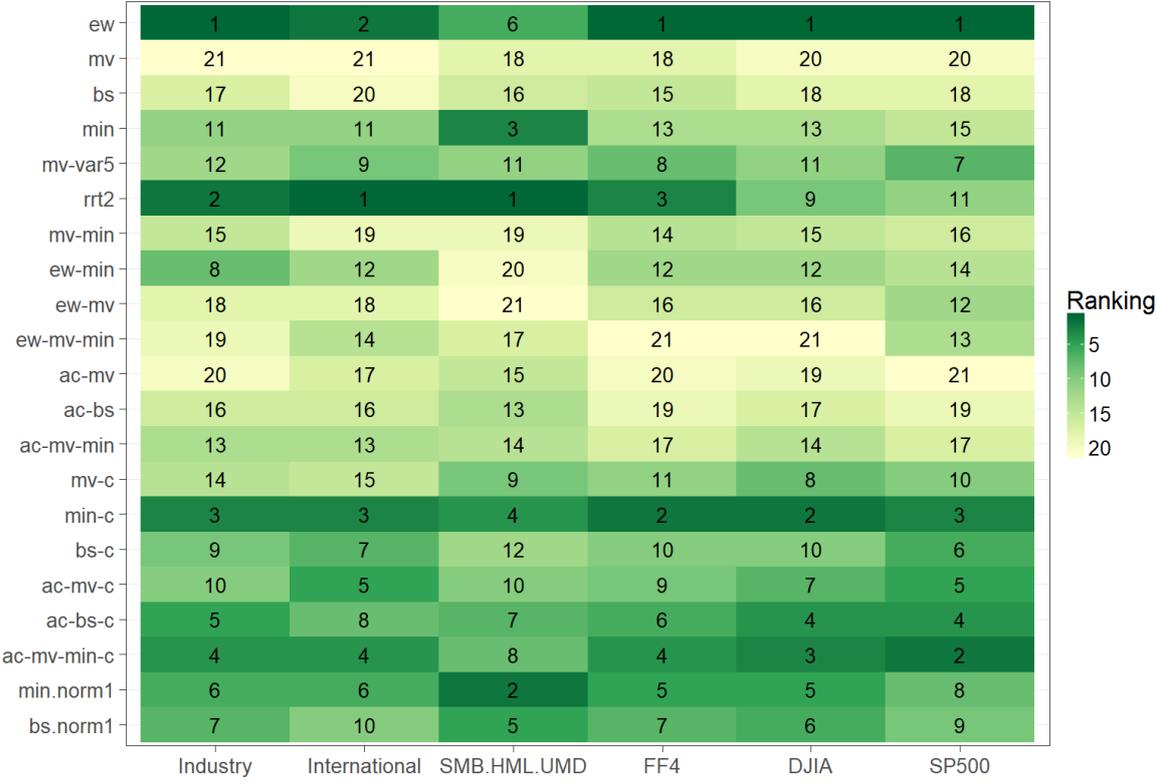

**Figure 3 | Average monthly turnover ranking**
This figure depicts the average monthly turnover rankings for the various portfolio selection strategies. For each dataset (x-axis) the portfolio rules (y-axis) are ranked from lowest (*dark green*) to highest (*bright yellow*) average monthly turnover.

## C.6   Return-loss relative to 1/N

**Table 8 | Monthly return-loss relative to 1/N**
For each empirical dataset this table reports the monthly return-loss relative to the 1/N rule, which is the extra return needed to yield the same Sharpe ratio as the benchmark and cover proportional transaction costs of fifty basis points.

| Strategy | Industry $N=10+1$ | International $N=8+1$ | SMB/HML/UMD $N=3+1$ | FF-4 $N=25+4$ | DJIA $N=33+4$ | SP500 $N=60+4$ |
|---|---|---|---|---|---|---|
| mv | 0.0035 | -0.0074 | 0.0013 | -0.0178 | 0.0187 | 0.0117 |
| bs | -0.0004 | -0.0033 | 0.0011 | -0.0149 | 0.0076 | 0.0025 |
| min | -0.0020 | 0.0005 | 0.0001 | -0.0070 | 0.0005 | -0.0015 |
| mv-var ($\delta = 5\%$) | -0.0013 | 0.0006 | 0.0000 | -0.0019 | 0.0001 | -0.0024 |
| rrt ($\omega = 2$) | 0.0013 | -0.0043 | 0.0012 | -0.0001 | 0.0001 | -0.0017 |
| mv-min | -0.0013 | -0.0018 | -0.0020 | -0.0128 | 0.0024 | -0.0014 |
| ew-min | -0.0020 | 0.0008 | 0.0152 | -0.0065 | 0.0003 | -0.0015 |
| ew-mv | 0.0012 | -0.0016 | 0.0136 | -0.0152 | 0.0041 | 0.0008 |
| ew-mv-min | -0.0010 | -0.0005 | -0.0009 | 0.4764 | 0.0233 | -0.0017 |
| ac-mv | 0.0025 | -0.0021 | -0.0023 | -0.0221 | 0.0146 | 0.0505 |
| ac-bs | -0.0007 | -0.0005 | -0.0012 | -0.0182 | 0.0059 | 0.0074 |
| ac-mv-min | -0.0014 | 0.0000 | -0.0013 | -0.0157 | 0.0016 | -0.0010 |
| mv-c | 0.0015 | -0.0012 | -0.0021 | -0.0003 | 0.0022 | 0.0045 |
| min-c | -0.0011 | 0.0006 | 0.0001 | -0.0016 | 0.0001 | -0.0021 |
| bs-c | -0.0002 | -0.0009 | -0.0016 | -0.0007 | 0.0005 | 0.0023 |
| ac-mv-c | 0.0000 | -0.0009 | -0.0016 | -0.0010 | -0.0002 | 0.0021 |
| ac-bs-c | -0.0006 | 0.0003 | -0.0010 | -0.0012 | -0.0003 | -0.0006 |
| ac-mv-min-c | -0.0009 | 0.0008 | -0.0011 | -0.0011 | 0.0000 | -0.0021 |
| min-norm ($\delta = 10\%$) | -0.0013 | 0.0005 | 0.0001 | -0.0020 | 0.0002 | -0.0023 |
| bs-norm ($\delta = 10\%$) | -0.0010 | 0.0008 | -0.0001 | -0.0019 | 0.0000 | -0.0022 |

## C.7 Return-loss relative to minimum-variance

**Table 9 | Monthly return-loss relative to minimum-variance**

For each empirical dataset listed in Table 2 this table reports the monthly return-loss relative to the minimum-variance portfolio, which is the extra return needed to yield the same Sharpe ratio as the benchmark and cover proportional transaction costs of fifty basis points.

| Strategy | Industry<br>*N=10+1* | International<br>*N=8+1* | SMB/HML/UMD<br>*N=3+1* | FF-4<br>*N=25+4* | DJIA<br>*N=33+4* | SP500<br>*N=60+4* |
|---|---|---|---|---|---|---|
| ew | 0.0023 | -0.0007 | -0.0001 | 0.0093 | -0.0006 | 0.0018 |
| mv | 0.0071 | -0.0102 | 0.0009 | -0.0040 | 0.0172 | 0.0160 |
| bs | 0.0019 | -0.0043 | 0.0008 | -0.0046 | 0.0068 | 0.0049 |
| mv-var ($\delta = 5\%$) | 0.0007 | 0.0000 | -0.0001 | 0.0060 | -0.0004 | -0.0011 |
| rrt ($\omega = 2$) | 0.0039 | -0.0049 | 0.0010 | 0.0095 | -0.0005 | -0.0001 |
| mv-min | 0.0008 | -0.0025 | -0.0022 | -0.0040 | 0.0018 | 0.0001 |
| ew-min | 0.0000 | 0.0002 | 0.0140 | 0.0004 | -0.0001 | 0.0000 |
| ew-mv | 0.0037 | -0.0023 | 0.0017 | -0.0038 | 0.0034 | 0.0026 |
| ew-mv-min | 0.0013 | -0.0012 | -0.0010 | 0.9122 | 0.0209 | -0.0003 |
| ac-mv | 0.0056 | -0.0027 | -0.0026 | 0.0021 | 0.0133 | 0.0599 |
| ac-bs | 0.0016 | -0.0010 | -0.0014 | 0.0004 | 0.0051 | 0.0103 |
| ac-mv-min | 0.0007 | -0.0005 | -0.0015 | 0.0007 | 0.0011 | 0.0005 |
| mv-c | 0.0039 | -0.0017 | -0.0023 | 0.0094 | 0.0014 | 0.0066 |
| min-c | 0.0009 | 0.0000 | 0.0000 | 0.0064 | -0.0004 | -0.0008 |
| bs-c | 0.0019 | -0.0014 | -0.0018 | 0.0087 | 0.0000 | 0.0039 |
| ac-mv-c | 0.0021 | -0.0014 | -0.0018 | 0.0080 | -0.0008 | 0.0038 |
| ac-bs-c | 0.0014 | -0.0002 | -0.0012 | 0.0075 | -0.0008 | 0.0008 |
| ac-mv-min-c | 0.0011 | 0.0003 | -0.0013 | 0.0073 | -0.0005 | -0.0008 |
| min-norm ($\delta = 10\%$) | 0.0007 | -0.0001 | 0.0000 | 0.0059 | -0.0003 | -0.0010 |
| bs-norm ($\delta = 10\%$) | 0.0009 | 0.0003 | -0.0002 | 0.0060 | -0.0004 | -0.0009 |